\documentclass[
    ,final            
]
  {aipproc}

\layoutstyle{8x11single}

\newcommand{\be}{\begin{equation}}
\newcommand{\ee}{\end{equation}}
\newcommand{\bea}{\begin{eqnarray}}
\newcommand{\eea}{\end{eqnarray}}

\def\lsim{\mathrel{\rlap{\lower4pt\hbox{\hskip1pt$\sim$}}\raise1pt\hbox{$<$}}}
\def\gsim{\mathrel{\rlap{\lower4pt\hbox{\hskip1pt$\sim$}}\raise1pt\hbox{$>$}}}
\def\nostrocostruttino#1\over#2{\mathrel{\mathop{\kern 0pt \rlap
{\hbox{$#1$}}} \hbox{\kern-.135em $#2$}}}

\def\kt{k_\perp}

\def\pp{p_\perp}

\def\kt{k_\perp}
\def\ktm2{k^2_{\perp{\rm max}}}

\begin{document}

\title{Transverse Momentum Dependent \\Partonic Distribution and Fragmentation Functions}




\author{M. Boglione}{
  address={Dipartimento di Fisica Teorica, Universit\`a di Torino, Via P.~Giuria 1, I-10125 Torino, Italy\\
              INFN, Sezione di Torino, Via P.~Giuria 1, I-10125 Torino, Italy}
}

%
%
%

\begin{abstract}
Our knowledge on the three-dimensional momentum structure of the hadrons
is encoded in the Transverse Momentum Dependent partonic distribution
and fragmentation functions (TMDs). A brief and updated review of the TMDs
and of the processes in which they might play a role is presented.

\end{abstract}

\maketitle


\section{Introduction}

The simple, collinear picture of fast moving nucleons as composite objects, made of constituent 
partons moving in the same direction of the parent nucleon and carrying a fraction $x$ 
of its lightcone momentum, has led to the effective description of the nucleon structure in terms 
of Parton Distribution Functions (PDFs). 
Recently, however, many experimental evidences -- 
like the $P_T$ distribution of hadrons produced in Semi-Inclusive Deep Inelastic Scattering (SIDIS) 
and the $q_T$ distribution of lepton pairs in Drell-Yan (DY) processes --  
have shown  the necessity of extending the investigation of the partonic structure of hadrons beyond 
the PDF picture, by exploring the parton motion and its spatial distribution in the transverse direction, 
perpendicular to the parent hadron momentum.  
The nucleon is indeed a 3-D object, which we need to study in all of its  three dimensions, 
longitudinal and transverse. Transverse Momentum Dependent distribution functions (TMDs) are the 3-D generalization 
of PDFs that allow us to do just that, as we shell see in what follows.

In the collinear approximation, three independent PDFs are sufficient to describe  
the proton structure: in addition to the unpolarized distribution function, $f_1(x)$, the helicity distribution 
function, $g_1(x)$, is connected to the longitudinal degrees of freedom, while the transversity PDF, $h_1(x)$, 
relates to the transverse spin structure of nucleons, 
and is the most important (and incidentally the most unknown) collinear PDF we are going to examine in this talk. 

\begin{figure}
  \includegraphics[height=.16\textheight]{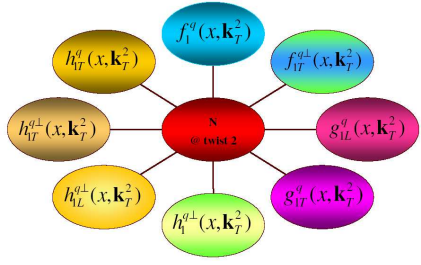}
  \caption{\label{TMDs}Leading twist transverse momentum dependent distribution functions. Courtesy of Aram Kotzinian.}
\end{figure}

In the TMD framework, to leading twist, the number of distribution functions raises from three 
(unpolarized, helicity and transversity) to eight, shown in Fig.~\ref{TMDs}: the unpolarized and helicity TMDs, 
$f_1(x,\kt)$ and $g_{1L}(x,\kt)$, are the 
straightforward generalization of the collinear unpolarized and helicity PDFs, defined in such a way that 
$f_1(x) = \int d^2 {\textit{\textbf{k}}} _\perp\; f_1(x,\kt)$ and  
$g_1(x) = \int d^2 {\textit{\textbf{k}}} _\perp \;g_{1L}(x,\kt)$, respectively.
Additional information on the transverse degrees of freedom of nucleon and/or partons is embedded in genuinely 
new TMDs, which do not survive in the collinear limit: this is the case of the Sivers and Boer-Mulders functions,  
$f_{1T}^\perp(x,\kt)$ and $h_1^\perp(x,\kt)$, which are related to the density number of unpolarized partons 
inside a transversely polarized proton and to the density number of transversely polarized partons inside 
an unpolarized proton, respectively. Similarly, two independent TMDs describe longitudinally polarized partons in a 
transversely polarized proton, $g_{1T}(x,\kt)$, and transversely polarized partons inside a longitudinally 
polarized proton, $h_{1L}^\perp(x,\kt)$: these functions are sometimes referred to as worm-gear TMDs.
Finally, the distribution of transversely polarized partons inside a transversely polarized proton is now embedded 
into two TMDs, $h_{1T}(x,\kt)$ and $h_{1T}^\perp(x,\kt)$, which suitably combined and integrated over 
${\textit{\textbf{k}}} _\perp$ give 
back the transversity function 
\(h_1(x)=\int d^2  {\textit{\textbf{k}}} _\perp \;[h_{1T}(x,\kt) + \frac{\kt^2}{2M^2}\, h_{1T}^\perp(x,\kt)]\;,\)
with $M$ the reference hadronic mass.

Like the collinear PDFs, the TMDs depend on the lightcone fraction $x$ and on the momentum transfer $Q^2$; 
additionally, they are functions of the parton intrinsic transverse momentum, $\kt$. 
Consequently, phenomenological studies of TMDs 
will inevitably face us with a number of open questions: first of all, the form of the $\kt$ distribution of TMDs. 
At present, it is commonly described in terms of simple models, the most widely used being 
a Gaussian $\kt$ distribution, completely factorized from the $x$ (and $Q^2$) dependence. This anzatz is supported by 
experimental evidences~\cite{:2008rv,Mkrtchyan:2007sr}, some phenomenological studies~\cite{Schweitzer:2010tt} and 
lattice calculations~\cite{Musch:2010ka}. However, we should mention that some important TMD models, quark-diquark or 
lightcone wave function models for example, prescribe a different and more complex $\kt$ dependence of TMDs, which does 
not show any factorization between the $x$ and $\kt$ behaviour, nor Gaussian $\kt$ tails.
Another very important issue regarding TMDs is their $Q^2$ evolution.
While QCD evolution equations predict the PDF dependence on the $Q^2$ scale, TMD evolution is a very complicated issue 
which has to be dealt with great care, starting from an appropriately devised definition of TMDs~\cite{Collins:2011}. 
Having done that, TMD evolution equations can then be derived~\cite{Aybat:2011zv}. 
Although this is a new field, the intention is to bring it soon to a stage in which numerical routines and grids for 
the computation of TMD evolution will be made available for public use, similarly to what has been done for PDF evolution.

\section{Phenomenological extraction of TMD functions}

The unpolarized and helicity PDFs have historically been extracted from global fits of Deep Inelastic Scattering 
(DIS) world-data, with excellent results in both cases, with the exception of the gluon and strange helicity 
functions, which still remain largely undetermined in the low-$x$ region. 

\begin{figure}
  \includegraphics[height=.18\textheight]{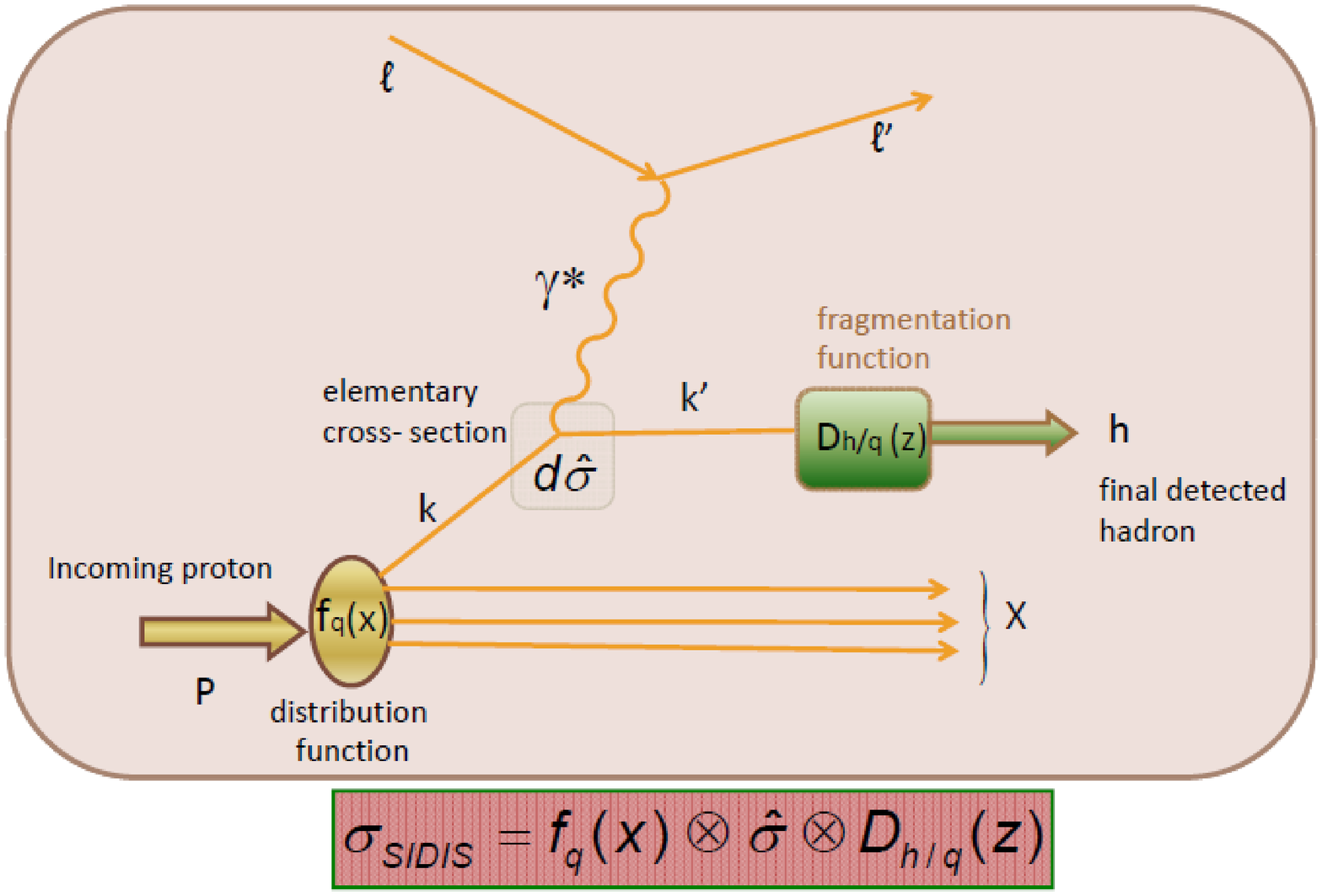}
\hspace*{2cm}
  \includegraphics[height=.18\textheight]{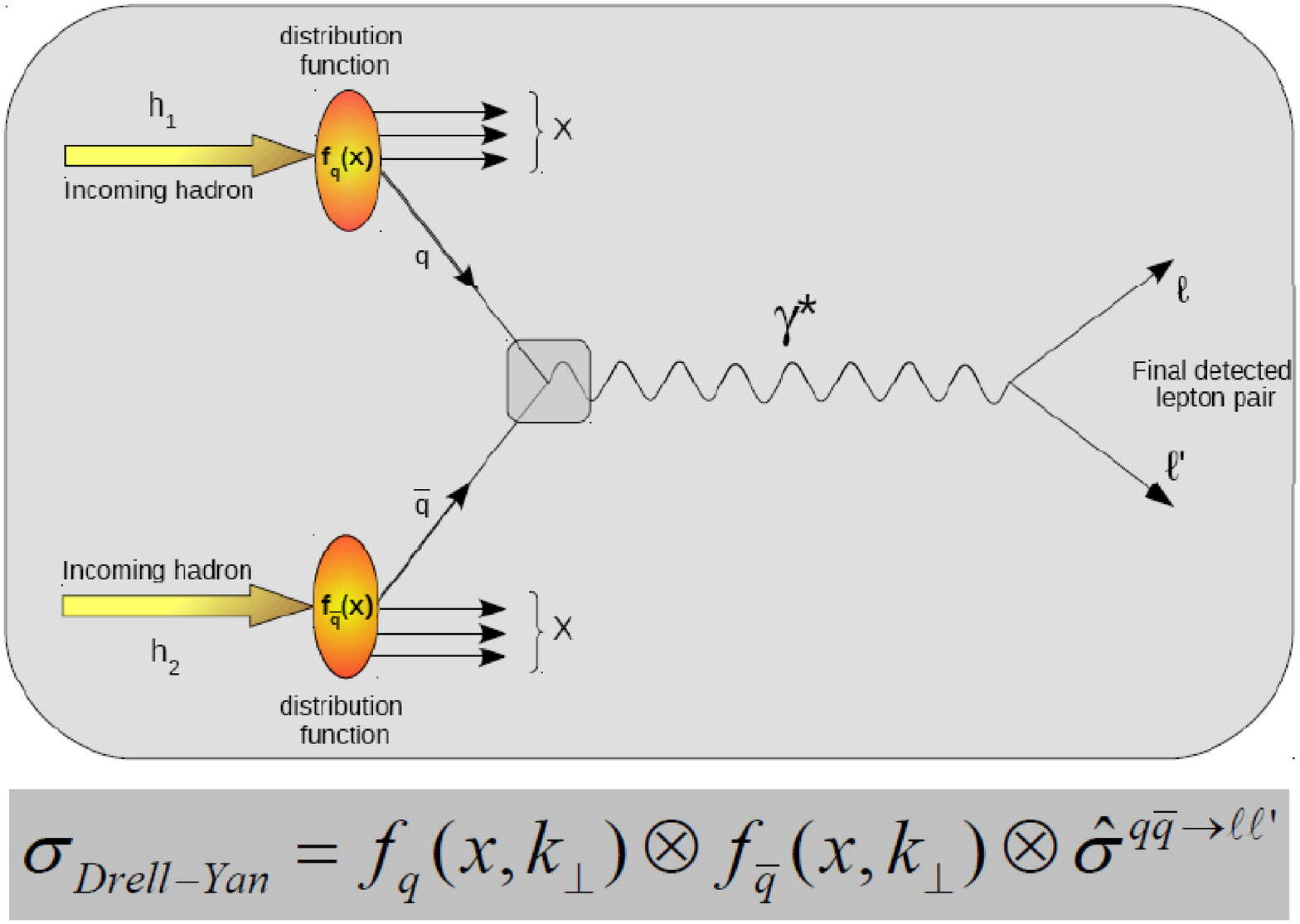}
  \caption{\label{sidis-dy} Pictorial illustration of a SIDIS process (left panel) 
                                  and of a DY scattering (right panel).}
\end{figure}

Due to its chiral-odd nature, the transversity distribution function cannot be measured in DIS processes, where 
the required helicity flip is forbidden by helicity conservation of the quark-photon vertex. 
The extraction of transversity demands more composite processes, for example Semi-Inclusive Deep Inelastic Scattering 
(SIDIS), in which the detection of a final hadron resulting from the fragmentation of the scattered parton 
is required, or DY processes, where the scattering of a quark-antiquark pair (generated from hadron-hadron 
scattering) results in the production of a lepton pair, see Fig.~\ref{sidis-dy}. 
In SIDIS, the chiral odd transversity distribution function appears coupled to a chiral odd fragmentation functions, 
the Collins function $H_1^\perp(z,\pp)$, describing the fragmentation of polarized partons into a spinless hadron:
\[
A_{UT}^{\sin(\phi_h+\phi_S)} \propto \sum _q h_1^q(x,\kt) \otimes d\Delta \hat{\sigma}^{\ell q \to \ell q} 
\otimes H_1^{q\perp}(z,\pp) \,.
\]
The double transverse spin asymmetries in DY processes, instead, are given by the convolution of two transversity distribution functions (even in the collinear framework), related to the
partons inside the initial proton and
antiproton:
\[
A_{TT} \propto \sum _q h_1^q (x_q,k_{\perp q}) 
\otimes h_1^{\bar{q}}(x_{\bar{q}},k _{\perp \bar{q}}) 
\otimes d\Delta \hat{\sigma}^{q {\bar q}  \to \ell ^+\ell^-} \,.
\]
In both cases the convolution of two chiral odd 
functions ensures a chiral even observable cross section.

While  doubly polarized proton-antiproton DY processes would be the golden channel to measure transversity, 
this faces us with the formidable challenge of building a beam of polarized antiprotons: this is presently beeing 
considered by the PAX collaboration~\cite{Barone:2005pu}, but no conclusive results on its feasibility are yet available. 
Proton-proton DY scattering processes, like those which can be measured at RHIC~\cite{ANDY-RHIC:2011} 
or Fermilab~\cite{Peng:2011ds} experiments, 
have the disadvantage of a tiny elementary cross section, suppressed by the 
antiquark distribution functions (antiquarks are sea partons in the proton), 
while proton-pion DY scattering processes, that the COMPASS Collaboration is planning to 
measure~\cite{Quintans:2011zz}, are not affected by this problem (as $\bar u$ and $\bar d$ antiquarks are 
valence partons in the pion), but they would require the detailed knowledge of the pion distribution functions 
(transversity PDF or Boer-Mulders TMD),  which is not presently available.

Beside transversity, DY processes can offer valuable information on various TMDs: for example, 
the Sivers function can be studied in single polarized DY 
scattering
~\cite{Anselmino:2002pd,Anselmino:2009st,Boglione:2011zw} 
\[
A_{N} \propto \sum _q f_{1T}^{\perp q} (x_q,k_{\perp q}) 
\otimes f_1^{\bar{q}}(x_{\bar{q}},k _{\perp \bar{q}}) 
\otimes \hat{\sigma}^{q {\bar q}  \to \ell ^+\ell^-} \,,
\]
while the Boer-Mulders function 
can be extracted from the unpolarized DY cross section~\cite{Barone:2010gk} 
\[
\sigma _{DY} \propto \sum _q h_1^{\perp q} (x_q,k_{\perp q}) 
\otimes h_1^{\perp \bar{q}}(x_{\bar{q}},k _{\perp \bar{q}}) 
\otimes \Delta \hat{\sigma}^{q {\bar q}  \to \ell ^+\ell^-} \,.
\]

At present, the best working ground to learn about TMD distribution and fragmentation functions is provided by 
SIDIS experiments, because of their large production rates and because the most general polarized cross section can 
easily be separated into individual contributions, each tagged by its own individual azimuthal dependence. 
Azimuthal moments, which isolate those terms one 
at a time, have been classified~\cite{Bacchetta:2006tn,Anselmino:2011ch}, and are widely used by the experimental 
collaborations to analyse their data.  

Among the TMDs which vanish in the collinear limit, the best known is the Sivers function, which was
phenomenologically extracted by exploiting various measurements of SIDIS azimuthal moments by the HERMES~\cite{:2009ti} and 
COMPASS~\cite{:2008dn} Collaborations:
\[
A_{UT}^{\sin(\phi_h-\phi_S)} \propto \sum _q f_{1T}^{\perp q}(x,\kt) \otimes \hat{\sigma}^{\ell q \to \ell q} 
\otimes D_1^{q}(z,\pp) \,.
\]
New and more sophisticated studies are now on the way which, beside including the most 
recent data, provide a more refined analysis of the uncertainties affecting the corresponding predictions.
Although all SIDIS measurements confirm the azimuthal dependence introduced by the intrinsic transverse motion, 
experimental data are still affected by some slight inconsistency which needs to be further clarified. 
The Sivers effect, for example, has now been studied by four independent experiments:
\begin{itemize}
\item HERMES (SIDIS on proton target~\cite{:2009ti}), which offers clear evidence of a non-zero Sivers effect;
\item COMPASS (SIDIS on deuteron target~\cite{:2008dn}), which finds a Sivers azimuthal moment compatible with zero,  
but crucial to fix the $d$ quark contribution to the Sivers function;
\item COMPASS (SIDIS on proton target~\cite{Alekseev:2010rw}), that seems to confirm a non zero Sivers effect 
for positive hadrons, although smaller than that measured by HERMES, but is affected by
a 10\% systematic error;
\item JLab Hall A (SIDIS on neutron target~\cite{Qian:2011py}) preliminary data, which provide a wider kinematical 
coverage compared to HERMES and COMPASS but are still affected by quite poor statistics.
\end{itemize}

Precise measurements of the Sivers function in both SIDIS and DY, are crucial to test another important aspect 
of the TMD approach.  In fact, while PDFs are universal, TMDs could in principle be process dependent. 
It has been shown~\cite{Collins:2002kn,Boer:2003cm}, that gauge links 
have to be included in the definition of TMDs to ensure their gauge invariance, and different gauge links correspond 
to either initial- or final-state interactions. This implies, for example, that the Sivers function extracted from 
SIDIS will have the opposite sign with respect to the Sivers function extracted from DY, due to the different 
signs of the two corresponding gauge links. Experimental confirmation of this prediction will indeed be a crucial 
test of our understanding of TMD factorization. Moreover, as the QCD predictive power is mainly based on the 
universality properties of distribution and fragmentation functions, this issues need to be put on a very firm ground 
as soon as possible. Notice that the Collins fragmentation function, instead, is believed to be universal~\cite{Yuan:2009dw}.

The Boer-Mulders function (which is chiral odd) is best examined by considering the unpolarized SIDIS cross section, 
where it appears coupled to the Collins fragmentation function in two different azimuthal moments:
\[
\langle\cos\phi_h\rangle \propto \frac{\kt}{Q}[A\,f_1^{ q} (x,\kt) \otimes D_1^{ q} (z,\pp) + B\;h_1^{\perp q} (x,\kt) 
\otimes H_1^{\perp q} (z,\pp)]\,,
\]
\[
\langle\cos2\phi_h\rangle \propto h_1^{\perp q} (x,\kt) \otimes H_1^{\perp q} (z,\pp)\,.
\]
Large experimental inconsistencies between HERMES and COMPASS results, in addition to possible sizeable higher-twist 
contaminations, make this analysis extremely complex. Interesting recent results can be found in 
Ref.~\cite{Barone:2009hw}, where a simple parametrization of the Boer-Mulders function in terms 
of the Sivers function was assumed and the Collins function was fixed to that extracted in 
Ref.~\cite{Anselmino:2008jk}. Moreover, in  Ref.~\cite{Boglione:2011wm}, sizeable effects due to the large $\kt$ 
Gaussian tails were taken into account by setting some kinematical upper limits to the $\kt$ integration, 
obtaining an unprecedented agreement between the model predictions and the $\langle\cos\phi_h\rangle$ experimental data.  

While waiting for new and high statistics DY data, a first extraction of 
transversity~\cite{Anselmino:2007fs,Anselmino:2008jk} has been achieved 
by simultaneously fitting SIDIS data provided by the COMPASS~\cite{Ageev:2006da} 
and HERMES~\cite{Airapetian:2004tw} collaborations and the BELLE measurements~\cite{Seidl:2008xc}  
of $e^+ e^- \to h_1 h_2$ reactions, which give access to the convolution of two Collins functions, describing the independent 
fragmentation of a quark and an antiquark into the two final detected hadrons.
\[A_{12}\propto H_1^{\perp q}(z_1,p_{\perp 1}) \otimes H_1^{\perp q}(z_2,p_{\perp 2}).\]
This is a delicate procedure, which involves two basic simultaneous steps: the determination of the Collins function at 
$Q^2 = 100$ GeV$^2$ essentially driven by the BELLE data, and the extraction of the transversity function from SIDIS data, 
where it appears coupled to the Collins function at $Q^2=2.5$ GeV$^2$. The evolution issue is therefore very important, 
and an updated analysis of the latest data that takes into account the issue of TMD evolution is on the way.

At this stage it is important to spend a few words on TMD factorization. As far as SIDIS and DY processes 
are concerned, two momentum scales exist: one large scale, $Q$, 
which ensures the feasibility of a perturbative expansion, and 
one small scale, the transverse momentum $P_T$ of the final hadron for SIDIS and transverse momentum $q_T$ of the 
final lepton pair for DY, to guarantee the sensitivity of 
the observables to the partonic transverse motion, such that  
$Q \gg P_T \simeq \kt \simeq \Lambda _{QCD}$ and  $Q \gg q_T$. For these processes, together with the two-hadron 
momentum imbalance in $e^+e^-$, TMD factorization has been proven to hold~\cite{Ji:2004wu,Ji:2004xq,Collins:2004nx}.

TMDs can, in principle, be studied in other processes, like single inclusive hadron production from polarized 
hadron-hadron scattering, $A^\uparrow + B \to C + X$, which produce surprisingly large single spin asymmetries, 
as measured by E704 at Fermilab~\cite{Adams:1991cs,Adams:1995gg} and, much more recently, by STAR~\cite{Eun:2011zz}, BRAHMS~\cite{Lee:2009ck} and PHENIX~\cite{Sarsour:2011zz} at RHIC. 
However, for processes where only one (large) scale exists, in this case the final hadronic $P_T$, 
TMD factorization has not been proven, and some attempts have been made to show that 
it can actually fail~\cite{Vogelsang:2007jk,Collins:2007nk,Rogers:2010dm}. 

Different strategies can therefore be used to analyze these processes: one possibility is to step back to 
the collinear approximation, where factorization can be safely applied, and the hadronic structure 
is embedded in the so-called higher-twist correlation functions, related to the strenght of color Lorentz force 
inside a spinning proton. First attempts have been performed to study the single spin asymmetries 
(SSAs) generated in polarized proton-proton single inclusive production of pions; the related higher twist 
correlation functions are related to the Sivers distribution,  but seem to be inconsistent in
sign with the Sivers function extracted from SIDIS data~\cite{Kang:2011hk}. 
Instead, one could assume TMD factorization to work and investigate what results can be obtained by simply 
applying the TMDs extracted from SIDIS to predict the SSAs measured in $p^\uparrow p \to \pi X$: this was done 
for example in Refs.~\cite{Boglione:2007dm,Anselmino:2009hk} with remarkably good results. 
Indeed these issues need further study and more 
refined analysis have to be performed to clarify all the open questions which remain unsolved. 

\section{Conclusions}

The exploration of the 3-D structure of the nucleons is a fundamental but formidable task, an extremely 
interesting challenge which needs further pursuing. Beside a well devised and efficient theoretical background, 
we need as many experimental data as possible, on as many different processes as possible,
to develop a satisfactory phenomenological picture of the nucleon phase space.

TMDs are instrumental to the study of the transverse
structure of nucleons, as they contain all the non-perturbative information on the transverse motion of partons and 
embed the correlation between spin and hadronic transverse momenta: presently, their knowledge covers a
limited kinematical range, which seriously limits the validity of extractions through data fits. 
Moreover, new fitting technologies should be introduced to limit the parametrization-dependence of the TMD fits, 
similarly to what happens for the PDFs.


\begin{theacknowledgments}
I thank M.~Anselmino, U.~D'Alesio, A.~Kotzinian, S.~Melis, F.~Murgia and A.~Prokudin  for our long and fruitful 
collaboration, and for their constant support throughout my work.  
\end{theacknowledgments}

\bibliographystyle{aipproc}  

\bibliography{newsample}

\IfFileExists{\jobname.bbl}{}
 {\typeout{}
  \typeout{******************************************}
  \typeout{** Please run "bibtex \jobname" to optain}
  \typeout{** the bibliography and then re-run LaTeX}
  \typeout{** twice to fix the references!}
  \typeout{******************************************}
  \typeout{}
 }

\end{document}